\documentstyle[12pt]{article}

\begin{document}

\title
{\Large {\bf A Spin - 3/2 Ising Model on a Square Lattice }}

\author
{N. Sh. Izmailian\thanks{permanent address: Department of Theoretical
Physics, Yerevan Physics Institute, Alikhanian Br.2, 375036 Yerevan,
Armenia} \\
\normalsize Bogoliubov Laboratory of Theoretical Physics,  \\
\normalsize Joint Institute for Nuclear Research, \\
\normalsize 141980, Dubna, Russia. }

\maketitle

\begin{abstract}
The spin - 3/2 Ising model described by the most general Hamiltonian with an
up-down symmetry
$$
-{\beta}H = \sum_{<ij>} \{ JS_iS_j + KS_i^2S_j^2 + LS_i^3S_j^3 + {M \over 2}
(S_iS_j^3 + S_jS_i^3) \} - \Delta \sum_{i} S_i^2
$$
is investigated on a square lattice. It is shown that this
model is reducible to an eight - vertex model on a surface in the parameter
space spanned by coupling constants $J$, $K$, $L$ and $M$. It is shown that
this model is equivalent to an exactly solvable free fermion model along two
lines in the parameter space. Consequently, the critical behaviour and, in
particular, the critical temperature for the second-order phase transitions
of the model was found exactly.
\end{abstract}

PACS number: 05.50+q, 75.10.Hk, 75.40.Cx.

\newpage
The Ising model has been one of the most actively studied systems. In
particular, spin - 1 and spin - 3/2 Ising models, which present a rich variety
of critical and multicritical phenomena, are of special interest.

The spin - 1 Ising model with bilinear ($J$) and biquadratic ($K$) nearest-
neighbor interactions and a single-ion potential ($\Delta$) is know as the
Blume-Emery-Griffiths (BEG) model \cite{A1}. The spin - 3/2 Ising model was
introduced to explain phase transitions in $DyVO_4$ and its phase diagrams
were obtained within the mean-field approximation \cite{A2}. Later, this model
was used in a study of tricritical properties of a ternary fluid mixture
\cite{A3}. The complete phase diagram of the spin - 3/2 Ising model model with
$L = M = 0$ has  been fully analyzed with the use of two different approaches:
mean field and Monte-Carlo techniques \cite{A4}.

Most studies for two-dimensional lattice mentioned above, are performed for
the square lattice, very few exact results for spin - 1 and spin - 3/2 Ising
models have been carried out only for the trivalent lattices, such as the
honeycomb lattice. Recently Horiguchi \cite{A5}, Wu \cite{A6} and Rosengren
and H\"{a}ggkvist \cite{A7} using  different theoretical approaches have exactly
solved the BEG model for the honeycomb lattice in the subspace of interaction
constants $J$ and $K$
\begin{equation}
\label{R1}
\exp(K)\cosh J = 1.
\end{equation}
The same result was obtained also and for the Bethe lattice \cite{A8}.

Wu and Wu \cite{A9} and Kolesik and Samaj \cite{A10} have considered the BEG
model in an external magnetic field $(H)$ and obtained an exact critical line
for all values of $H$. Recently, Lipowski and Suzuki \cite{A11} and Ananikian
and Izmailian \cite{A12} found the conditions under which the spin - 3/2 Ising
model on the honeycomb lattice has the same partition function as the exactly
solvable zero-field spin - 1/2 Ising model. Very recently, Horiguchi \cite{A13}
proposed a general method by which a general spin-$S$ Ising model in expressed
in terms of an Ising model of spin $\pm 1$ and spin less then $S$. This exact
results provides an excellent tool for comparing the accuracy of different
approximation schemes mentioned above.

In the present paper, we have solved exactly the most general spin - 3/2 Ising
model for the two-dimensional square lattice in the subspace of the four-
dimentional space spanned by the coupling constants $J$, $K$, $L$ and $M$. We
show that this model is reducible to an eight-vertex model, for which the exact
solution is not known at the present except for a few special cases
\cite{A14,A15}. Moreover, we have  established the equivalence of our model on the square
lattice with one of the special cases, the so-called  "free fermion model",
which is the eight-vertex model under the "free fermion condition", along two
lines in the four-dimentional space spanned by the coupling constants $J$, $K$,
$L$ and $M$.

1. We consider the most general spin - 3/2 Ising model with an nearest-neighbor
interaction and an up-down symmetry, which is described by the following
Hamiltonian
\begin{equation}
\label{R2}
-{\beta}H = \sum_{<ij>} \{ JS_iS_j + KS_i^2S_j^2 + LS_i^3S_j^3 + {M \over 2}
(S_iS_j^3 + S_jS_i^3) \} - \Delta \sum_{i} S_i^2,
\end{equation}
where $S_i = \pm {1 \over 2}, \pm {3 \over 2} $ is the spin variable at site
$i$ and $<ij>$ indicates the summation over the pairs of nearest-neighbor
sites.

The equilibrium statistics of the system given by Eq.(\ref{R2}) is determined
by the partition function $Z = \sum_{\{ s \}} \exp ({-\beta} H)$, where $\beta=
1/k_B T$ is the inverse temperature and the sum is over all spin configurations.

It is not difficult to show (for details see Ananikian and Izmailian \cite{A12})
that if
\begin{equation}
\label{R3}
\cases{\tanh^2J_1 = \tanh J_2 \tanh J_0 \cr \exp(-4K) = \cosh (J_2 - J_0)},
\end{equation}
where
$$
J_0 = {1 \over 4} J + {1 \over 16} M + {1 \over 64} L, \quad
J_1 = {3 \over 4} J + {15 \over 16} M + {27 \over 64} L, \quad
J_2 = {9 \over 4} J + {81 \over 16} M + {729 \over 64} L,
$$
we can write the following identity
$$
\exp { \{ JS_1S_2 + KS_1^2S_2^2 + LS_1^3S_2^3 + {M \over 2}
(S_1S_2^3 + S_2S_1^3) \} } =
$$
\begin{equation}
\label{R4}
= \alpha_{0} \exp[R(S_1^2 + S_2^2 - {1 \over 2})] {\{1 + 4\tanh J_0S_1S_2\exp
[R_0(S_1^2 + S_2^2 - {1 \over 2})] \} },
\end{equation}
where $ \alpha_0 = \exp(K/16)\cosh J_0$ and
\begin{equation}
\label{R5}
\exp(2R) = \frac{\cosh J_1}{\cosh J_0} \exp({K \over 2}), \quad
\exp(4R_0) = \frac{\tanh J_2}{9\tanh J_0}.
\end{equation}

Then the partition function defined by Hamiltonian in Eq.(\ref{R2}) can be
written as:
\begin{equation}
\label{R6}
Z = {(\alpha_{0} e^{- {R \over 2}})}^E \sum_{ \{ s \} } \prod_{<ij>} {\{
1 + 4\tanh J_0S_iS_j \exp[R_0(S_i^2 + S_j^2 - {1 \over 2})] \} } \prod_{i}
\exp({-\Delta_0 S_i^2}),
\end{equation}
where $E$ is the total number of edges, $\Delta_0 = \Delta - \gamma R$, $\gamma$
is the coordination number of a lattice, and the first product in Eq.(\ref{R6})
is extended over all pairs of neighbouring sites. This result is valid for any
arbitrary lattice. It should be noted there that Eq.(\ref{R3}) is an analog of
the condition in Eq.(\ref{R1}) for the spin - 1 Ising model.

Thus, we obtain the condition given in Eq.(\ref{R3}), which defines the surface
in the space spanned by the coupling constants $J$, $K$, $L$ and $M$, where the
partition function is written in the form of Eq.(\ref{R6}). Recently, the
equivalence of the spin - 3/2 Ising model on the honeycomb lattice with a
zero-field spin - 1/2 Ising model on the same lattice in the subspace given in
Eq.(\ref{R3}) have been established \cite{A11,A12}.

2. In the present section, we investigate the most general spin - 3/2 Ising
model on the square lattice, described by the Hamiltonian in Eq.(\ref{R2}). We
show that this model is reducible to an  eight-vertex model and can be exactly
solved on the two nontrivial lines in the surface given by Eq.(\ref{R3}).

Now, consider a square lattice (where the coordination number $\gamma$ is equal
to four) composed of $N$ sites (or vertices) and of $2N$ lattice edges. Then we
expand the product $\prod_{<ij>}$ in Eq.(\ref{R6}) and represent graphically
each term in the expansion as follows: draw a  dotted or solid line over the
lattice edge ($ij$) if the corresponding term in the expansion contains the
factor 1 or $4\tanh J_0S_iS_j \exp[R_0(S_i^2 + S_j^2 - {1 \over 2})] $. This leads
to eight different kinds of configurations, shown in Fig.1, that can occur at a
vertex. We assign a Boltzman weight $\{ w_i \}$ to a vertex having $2n$ solid
and $4-2n$ broken edges. When, the summations for all sites are carried out, we
obtained the following values for weights to vertices:

\begin{eqnarray}
\label{R7}
\sum_{s_i = \pm {1 \over 2}, \pm {3 \over 2}}t^nS_i^n \exp[(nR_0 - \Delta_0)S_i^2] =
\left \{ \begin {array}{llll}
a & = 2 e^{- \frac{\Delta_0}{4}} [1 + e^{-2 \Delta_0}], & n = 0 \\
b & = 2t e^{- \frac{\Delta_0}{4}} [1 + 9e^{4R_0 - 2 \Delta_0}], & n = 2 \\
c & = 2t^2 e^{- \frac{\Delta_0}{4}} [1 + 81e^{8R_0 - 2 \Delta_0}], & n = 4 \\
& = 0, & n = 1, 3
\end{array}
\right.
\end{eqnarray}

where $t=\tanh J_0$ and $n$ is the number of lines with site $i$ as an end-point.

These facts enable us to rewrite Eq.(\ref{R6}) in the form where the sum is over
all line configurations on the square lattice having an even number of lines
into each site. For the square lattice, this leads to an eight-vertex model
shown in Fig.1 with the vertex weights

\begin{equation}
\label{R8}
w_1 = a, \quad w_2 = c \quad {\mbox and} \quad w_3 = \dots = w_8 = b,
\end{equation}
where $a$, $b$ and $c$ are given in Eq.(\ref{R7}).

Thus, we obtain the exact equivalence
\begin{equation}
\label{R9}
Z = (\alpha_0 e^{-{R_0 \over 2}})^{2N} Z_{8v}(\{w_i\}),
\end{equation}
where $Z_8v(\{w_i\})$ is the partition function of eight-vertex model given by
\begin{equation}
\label{R10}
Z_{8v}(\{w_i\}) = 2\sum_{\mbox{\tiny{all line configurations}}} \prod_i w_i.
\end{equation}

The factor of 2 comes from the fact that a reversing of all spins leaves the
line configurations unchanged.

The eight-vertex model on the square lattice plays an important role in the
study of phase transitions in lattice systems. Unfortunately, except in some
special cases \cite{A14,A15} the behavior of this general model is not known.
The exact expression of the free energy of this model was first obtained by
Fan and Wu \cite{A15} and by Baxter \cite{A14} in respective conditions. In
particular, the former authors solved the "free fermion model". This model is
defined as a particular case of the eight-vertex model in which the vertex
weights satisfy the relation:
\begin{equation}
\label{R11}
w_1w_2 + w_3w_4 = w_5w_6 + w_7w_8,
\end{equation}
which is called the "free fermion condition", as in the $S$ - matrix formulation
of the eight-vertex problem, this condition is equivalent to the consideration
of non-interacting many-fermion system \cite{A16}.

It is readily verified, using Eqs.(\ref{R7}) and (\ref{R8}), that the "free
fermion condition" in Eq.(\ref{R11}) is satisfied if, and only if
\begin{equation}
\label{R12}
\exp(4R_0) = {1 \over 9}.
\end{equation}

This equation together with Eqs.(\ref{R3}) and (\ref{R5}) give us the following
two nontrivial lines in the four-dimensional parameter space spanned by coupling
constants $J$, $K$, $L$ and $M$, i. e.,
\begin{equation}
\label{R13}
\cases{(i) \quad 16J = 49L = -14M, \quad K=0; \quad (J_0 = \frac{9}{49} J)\cr
(ii) \quad 16J = 169L = -26M, \quad K = 0; \quad (J_0 = \frac{36}{169} J)\cr},
\end{equation}
on which the spin - 3/2 Ising model is equivalent to exactly solvable free
fermion model. Here, we remark that except at the trivial point ($J = L = M =
K = 0$) for special ($M = 0$) spin - 3/2 Ising model, in general the exactly
solvable case can not be obtained \cite{A11}.

The equivalence, given explicity by Eqs.(\ref{R9}), (\ref{R10}) and (\ref{R13})
permits us to deduce exact analytic properties of the general spin - 3/2 Ising
model.

The condition given by Eq.(\ref{R13}) can be also interpretated in another way.
Consider, for example, the case (i):
Introducing the two new spin variables ${\sigma}_i$ and $t_i$
\begin{equation}
\label{R14}
{\sigma}_i = {4\over 3}S_i(S_i^2-{7\over 4}), \quad t_i = S_i^2-{5\over 4}
\end{equation}
we can write the Hamiltonian given by Eq.(\ref{R2}) in the form
$$
-{\beta}H = {9\over 16}L\sum_{<ij>} {\sigma}_i{\sigma}_j-\Delta\sum_{i} t_i-
{5\over 4}\Delta
$$
As it easy to see from Eq.(\ref{R14}) the two states $S_i = {3\over 2},
-{1\over 2}$ were transformed into ${\sigma}_i = 1$ and two states $S_i =
-{3\over 2},{1\over 2}$ into ${\sigma}_i = -1$, while the states $S_i =
\pm{3\over 2}$ and $S_i = \pm{1\over 2}$ into $t_i = 1$ and $t_i = -1$
respectively. This gives a one-to-one correspondence between $S_i$ and a pair
of spins $({\sigma}_i, t_i)$. Sowe can say that the our new spin variables
${\sigma}_i$ and $t_i$ are independent and consequently the summation in
$\sum_{(\sigma,t)}\exp(-{\beta}H)$ can be carry out separatly.

Now, we consider the thermodynamic properties of our model. A closed expression
for the free energy of the free fermion model is well known \cite{A15}  and
after some algebraic manipulation, we obtain, in the large $N$ limit, the free
energy for the spin - 3/2 Ising model on the square lattice in the subspace
given by Eq.(\ref{R13})

\begin{equation}
\label{R15}
-{\beta f} = \ln{\{2 \sqrt3 \exp(-{\frac{\Delta}{4}})[1 + \exp(-2{\Delta})]\}} +
{\frac{1}{8{\pi}^2}}\int_{0}^{2\pi} \ d{\vartheta} \int_{0}^{2\pi} \ d{\varphi}
\ln[c^2 + s(\cos{\vartheta} + \cos{\varphi})],
\end{equation}
where $c = \cosh 2J_0$, $s = \sinh 2J_0$ and $J_0$ is determined from
Eq.(\ref{R13}).

Since the second derivative of the free energy in Eq.(\ref{R15}) diverges
logarithmically at $s = s_c = 1$, the critical behavior of  our model are
summarized as follows: The spin - 3/2 Ising model exhibits a first-order phase
transition if $s > 1$, a second-order phase transition if $s = 1$ and no
transition at all if $s < 1$. Thus, a second-order phase transition occurs at
a temperature determined by $\sinh 2J_0 = 1$. This critical condition gives us
the two $\lambda$ - lines in the space spanned by $J$ and $\Delta$
$$
(i) \quad  J = {49 \over 18} \ln(1 + \sqrt2) = 3,0886... \quad and \quad
\Delta - arbitrary,
$$
$$
(ii) \quad  J = {169 \over 72} \ln(1 + \sqrt2) = 2,6631... \quad and \quad
\Delta - arbitrary,
$$
where the spin - 3/2 Ising model exhibits an Ising-type phase transition
(logarithmic specific heat singularity).

Besides the intrinsic interest surrounding the spin - $S$ Ising model and
possible applications in real physical situations, one is further attracted to
the search for  obtaining exact nontrivial solutions. So far soluble problems
are very few in numbers. In this paper, an exact solution for most general
spin - 3/2 Ising model is obtained. It is shown that this model is equivalent
to exactly solvable free fermion model along two lines in the parameter space
given by Eq.(\ref{R13}). In particular, the analytical expressions for the free
energy per spin the $\lambda$ - lines of Ising-type phase transition were found
exactly.

Finally, we point out, that our result suggests that equivalence of spin - 3/2
Ising model to the free fermion model can be extended to a most general Ising
model with an half-integer spin. That is, the existence of the exact solvable
case for the Ising model with an half-integer spin results from the absence of
the $S_i^z = 0$ state.

I wish to thank Professor G.-C. Wang for her hospitality while I was at the
Rensselaer Polytechnic Institute, where this work was carried out. I also
grateful to Professor G.-C. Wang for a  reading of the manuscript. This work
was supported in part by Soros Foundation grant awarded by the American Physical
Society, grant 211-5291 YPI of the German Bundasminsterium fur Forschung and
Technologie and CAST grant program of the National Academy of Sciences USA.

\end{document}